\documentclass[conference]{IEEEtran}
\IEEEoverridecommandlockouts
\usepackage{cite}
\usepackage{amsmath,amssymb,amsfonts,mathtools,bm}

\usepackage{amsthm}
\usepackage{graphicx}
\usepackage{booktabs}
\usepackage{xcolor}
\usepackage{algorithm}
\usepackage{algpseudocode}
\makeatletter
\renewcommand\fs@ruled{\def\@fs@cfont{\bfseries}\let\@fs@capt\floatc@ruled
  \def\@fs@pre{\kern6pt\hrule height.8pt depth0pt \kern2pt}%
  \def\@fs@post{\kern2pt\hrule\relax}%
  \def\@fs@mid{\kern2pt\hrule\kern2pt}%
  \let\@fs@iftopcapt\iftrue}
\makeatother
\floatstyle{ruled}\restylefloat{algorithm}
\usepackage{tikz}
\usetikzlibrary{arrows.meta}
\usepackage[font=footnotesize,labelsep=period,justification=justified,singlelinecheck=false]{caption}
\usepackage{dblfloatfix}
\usepackage[hidelinks]{hyperref}

\theoremstyle{plain}
\newtheorem{theorem}{Theorem}
\newtheorem{proposition}{Proposition}
\newtheorem{lemma}{Lemma}
\newtheorem{corollary}{Corollary}
\newtheoremstyle{remarkcolon}{3pt}{3pt}{\normalfont}{\parindent}{\bfseries}{:}{0.5em}{}
\theoremstyle{remarkcolon}
\newtheorem{remark}{Remark}

\newcommand{\E}{\mathbb{E}}
\newcommand{\C}{\mathcal{S}}
\newcommand{\CN}{\mathcal{CN}}
\newcommand{\mmse}{\mathsf{MMSE}}
\newcommand{\tr}{\operatorname{tr}}
\newcommand{\Cov}{\operatorname{cov}}
\newcommand{\dB}{\,\mathrm{dB}}
\newcommand{\rev}[1]{{\color{black}#1}}

\begin{document}

\title{When Does Inter-BS Interference Hurt Sensing in ISAC? An Exact Fisher Information Analysis under Finite-Alphabet Interference}

\author{\IEEEauthorblockN{Kawon Han}
\IEEEauthorblockA{\textit{UNIST}\\
Ulsan, South Korea\\
{kawon.han@unist.ac.kr}}
\and
\IEEEauthorblockN{Kaitao Meng}
\IEEEauthorblockA{\textit{The University of Manchester}\\
Manchester, United Kingdom\\
{kaitao.meng@manchester.ac.uk}}
\and
\IEEEauthorblockN{Fan Liu}
\IEEEauthorblockA{\textit{Southeast University}\\
Nanjing, China\\
{fan.liu@seu.edu.cn,}}
\and
\IEEEauthorblockN{Christos Masouros}
\IEEEauthorblockA{\textit{University College London}\\
London, United Kingdom\\
{c.masouros@ucl.ac.uk}}
}

\maketitle
\begin{abstract}
In networked integrated sensing and communication (ISAC), a base station (BS) performing sensing using its own downlink signals is subject to interference from the downlink transmissions of neighboring BSs. This paper investigates the impact of the signal constellation and waveform on the sensing performance in the presence of such inter-BS interference. For any linear receiver, we show that the mean-squared cross-ambiguity between two random ISAC signals is invariant to the modulation bases and constellations. We then derive the exact Fisher information under finite-alphabet interference with a flat-fading link, which equals the interference-free Fisher information scaled by $1-\gamma\,\mmse_{\C}(\gamma)$, where $\gamma$ denotes the interference-to-noise ratio (INR) and $\mmse_{\C}$ is the minimum mean-square error (MMSE) in estimating an interfering data symbol from its noise-corrupted observation at the sensing receiver. Consequently, Gaussian interference is the worst case and corresponds to the CRB loss of the matched filtering, whereas finite-alphabet interference induces a non-monotonic CRB loss that peaks at intermediate INR and vanishes at high INR. The loss is waveform-independent for the flat-fading BS-BS link, and an expectation-maximization (EM) interference-cancellation receiver approaches the theoretical bound for delay estimation.
\end{abstract}


\section{Introduction}
Integrated sensing and communication (ISAC) has attracted significant attention as a key enabler of next-generation wireless networks, in which the downlink (DL) data signals of cellular base stations (BSs) are reused for radar sensing~\cite{liu2025sensing,sturm2011}. In a dense, TDD-synchronized network, a BS that performs full-duplex monostatic sensing inevitably receives the DL transmissions of its neighbors on the same time-frequency resources \cite{han2025network, meng2025network}. Since BS-to-BS links are typically line-of-sight (LoS), this inter-BS interference can exceed the thermal noise by tens of decibels, and it fundamentally limits the sensing performance of network-level ISAC \cite{meng2024network}.

Since communication payloads are random, the sensing performance of a data-bearing signal depends on its modulation constellation and basis. OFDM achieves the lowest average ranging sidelobe under sub-Gaussian constellations~\cite{liu2025cp}, while the sidelobes induced by high-order QAM can be suppressed by mismatched filtering receiver~\cite{keskin2025fundamental}. Constellation shaping or selection balances this randomness against the communication rate~\cite{du2024reshaping,geiger2026constellation,meng2026constellation}, and the delay-estimation accuracy of OFDM-ISAC under matched filtering (MF) and reciprocal filtering (RF) has been characterized via the fourth and inverse second moments of the constellation~\cite{han2026ofdm}. However, all of these results concern the auto-ambiguity of a BS’s own signal.

For inter-BS (inter-cell) interference, constellation selection and power allocation in multi-cell OFDM-ISAC have been studied under MF and RF processing~\cite{meng2026constellation2}, and blind estimation under co-channel OFDM sources has also been investigated~\cite{pu2026blind}, which characterize the interference by its constellation moments and bound the performance with a Gaussian surrogate, respectively. On the communication side, it is well known that discrete interference \rev{causes less performance degradation} than Gaussian interference~\cite{dytso2016tin}. Despite these advances, an exact estimation-theoretic characterization of sensing under finite-alphabet inter-BS interference remains unexplored, particularly regarding how interference strength and modulation determine the achievable sensing accuracy.

To address this gap, we develop an estimation-theoretic framework for sensing under finite-alphabet inter-BS interference. First, we first show that the mean-squared cross-ambiguity is invariant to the modulation bases and constellations, so that no constellation or waveform design benefits a linear receiver. Then, we derive the exact Fisher information matrix (FIM) under finite-alphabet interference via a complex Brown identity, revealing that Gaussian interference is the worst case and that finite-alphabet interference yields a non-monotonic loss that vanishes at high INR. We further prove that the loss is waveform-independent for a flat-fading and synchronized BS-BS link, and show that an expectation-maximization (EM)-based soft interference-cancellation receiver \rev{approaches} the bound.

\section{System Model}\label{sec:model}

\begin{figure}[t]
\centering
\includegraphics[width=\columnwidth]{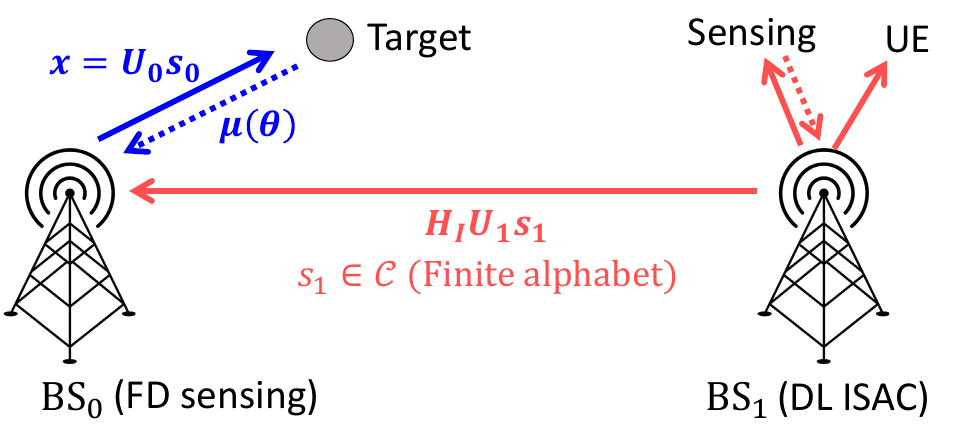}
\caption{ISAC under Inter-BS interference: BS$_0$ senses with its own DL data while receiving the DL data of BS$_1$ on the same time-frequency resources.}
\label{fig:sys}
\end{figure}

\subsection{Transmit Signal Model}
As illustrated in Fig.~\ref{fig:sys}, BS$_0$ performs full-duplex (FD) monostatic sensing with its DL data signal, while a neighboring BS$_1$ transmits independent DL data over the same time-frequency resources. To cover a broad class of waveforms, we consider a generic block transmission, in which each BS maps $N$ data symbols onto a length-$N$ block through a unitary modulation basis and appends a cyclic prefix (CP), i.e.,
\begin{equation}\label{eq:tx}
\bm x_0=\bm U_0\bm s_0,\qquad \bm x_1=\bm U_1\bm s_1,
\end{equation}
where $\bm s_0$ and $\bm s_1$ are the symbol vectors of BS$_0$ and BS$_1$, and $\bm U_0,\bm U_1\in\mathbb{C}^{N\times N}$ are unitary. This model includes single carrier (SC, $\bm U_1=\bm I$), CP-OFDM ($\bm U_1=\rev{\bm W_N^H}$ with the unitary DFT matrix $\rev{\bm W_N}$), and OTFS ($\bm U_1=\rev{\bm W_{N_{\mathrm D}}^H}\otimes\bm I_{N/N_{\mathrm D}}$ with $N_{\mathrm D}$ Doppler bins), as well as DFT-spread OFDM and AFDM. The symbols are i.i.d., zero-mean, and unit-power, and the interfering symbols are drawn from the constellation $\C$. We define the fourth moment (kurtosis) and the inverse second moment of a constellation as $\mu_4=\E[|s|^4]$ and $\nu_{-2}=\E[|s|^{-2}]$.

\subsection{Sensing Received Signal Model}
After self-interference suppression and CP removal, the received signal of BS$_0$ is
\begin{align}
    \bm r &=\bm\mu(\bm\eta)+\bm A\bm s_1+\bm n, \label{eq:model} \\
    \bm A &=\bm H_{\mathrm I}\bm U_1,\quad \bm n\sim\CN(\bm 0,\sigma^2\bm I),
\end{align}
where $\bm\mu(\bm\eta)$ is the target echo of BS$_0$'s own signal $\bm x_0$, e.g., $\bm\mu(\bm\eta)=\alpha\bm J_{\tau,\nu}\bm x_0$ for a single target with complex amplitude $\alpha$, delay $\tau$, Doppler $\nu$, and the delay-Doppler shift operator $\bm J_{\tau,\nu}$, and $\bm\eta$ collects the real-valued sensing parameters. $\bm H_{\mathrm I}$ is the circulant matrix of the BS$_1$-to-BS$_0$ channel $\bm h_{\mathrm I}$, and we define the per-sample INR as $\gamma=\|\bm h_{\mathrm I}\|^2/\sigma^2$. For a CP-OFDM interferer, $\bm A=\rev{\bm W_N^H}\mathrm{diag}(H_k)$, which yields the conventional per-subcarrier model $\tilde r_k=\tilde\mu_k(\bm\eta)+H_k\rev{s_{1,k}}+\tilde n_k$ after the DFT. In this study, we adopt the following assumptions:
\begin{itemize}
\item[(A1)] $\bm s_1$ has i.i.d.\ entries uniform on $\C$ and is independent of $\bm\eta$, $\bm s_0$, and $\bm n$.
\item[(A2)] The propagation delay of the BS$_1$ signal is within the CP, and the residual carrier frequency offset (CFO) is negligible.
\item[(A3)] BS$_0$ knows its own data, the static $\bm H_{\mathrm I}$ (estimated from the reference signals of BS$_1$), $\bm U_1$, and $\C$.
\item[(A4)] $\C$ is invariant to $90^\circ$ rotations, and thus $\E[s_1^2]=0$.
\end{itemize}

\section{Main Results}\label{sec:main}

\subsection{Linear Receivers: Second-Order Invariance}\label{sec:second}
With the unitary delay-Doppler shift $\bm J_{\ell,q}=\mathrm{diag}(e^{j2\pi qn/N})\bm\Pi^{\ell}$, where $\bm\Pi$ is the cyclic shift matrix, the periodic cross-ambiguity function between the signals in~\eqref{eq:tx} is defined as
\begin{align}
\chi(\ell,q)&=\bm x_1^H\bm J_{\ell,q}\bm x_0=\bm s_1^H\bm G\bm s_0, \label{eq:caf} \\
\bm G&=\bm U_1^H\bm J_{\ell,q}\bm U_0.
\end{align}
With this in hand, we first show that the second moment of $\chi$ is invariant to the modulation bases and constellations for any $N$, and so is its fourth moment for large $N$.

\begin{proposition}\label{prop:cross}
For all $(\ell,q)$, all unitary bases $\bm U_0,\bm U_1$, and all unit-power constellations with $\E[s_0^2]=\E[s_1^2]=0$, $\E[|\chi(\ell,q)|^2]=N$. Moreover, with the kurtosis $\mu_{4,s_0}$ and $\mu_{4,s_1}$ of $s_0$ and $s_1$,
\begin{multline}\label{eq:4th}
\E[|\chi|^4]=2N^2+2N(\mu_{4,s_0}+\mu_{4,s_1}-3)\\
+(\mu_{4,s_0}-2)(\mu_{4,s_1}-2)\|\bm G\|_4^4,
\end{multline}
where $\|\bm G\|_4^4=\sum_{i,j}|G_{ij}|^4\in[1,N]$. 
\end{proposition}
\begin{proof}
Please refer to Appendix~\ref{app:prop1}.
\end{proof}
The upper and lower bounds of $\|\bm G\|_4^4$ are attained iff $\bm G$ is monomial (e.g., OFDM-OFDM) and iff $|G_{ij}|^2=1/N$ (e.g., OFDM-SC), respectively. Since $\E[|\chi(\ell,q)|^2]=N$ for all $(\ell,q)$, the interference after MF forms a flat floor over the delay-Doppler map at approximately $1/N$ relative to the main lobe $\E[|\bm x_0^H\bm x_0|^2]=N^2+N(\mu_{4,\rev{s_0}}-1)$ regardless of the basis and constellation of the interferer unlike the auto-ambiguity sidelobes~\cite{liu2025cp}. Moreover,~\eqref{eq:4th} yields $\E[|\chi|^4]/(\E[|\chi|^2])^2=2+c/N$ with $c$ bounded in $N$, implying that the floor fluctuates as Gaussian noise for large $N$. 

\begin{remark}\label{rem:linear}
Linear filters such as MF, RF and LMMSE depend on the interference only through its covariance $\E[(\bm A\bm s_1)(\bm A\bm s_1)^H]=\bm H_{\mathrm I}\bm H_{\mathrm I}^H$, which is independent of $\bm U_1$ and $\C$, and also of $\bm s_0$ by (A1). Hence, this holds even when the linear filter is designed for each realization of the known $\bm s_0$. \rev{Therefore}, the \rev{choice of the waveform and constellation} of the interferer cannot affect linear receivers, and only the own constellation enters through $\nu_{-2}$ for RF processing~\cite{han2026ofdm,meng2026constellation2}. Exploiting the interference structure thus requires non-linear processing.
\end{remark}

\subsection{Fisher Information under Finite-Alphabet Interference}\label{sec:fisher}
Since $\bm\eta$ enters~\eqref{eq:model} only through $\bm\mu(\bm\eta)$, the FIM is governed by the Fisher information of $\bm w=\bm A\bm s_1+\bm n$ with respect to a location shift. For finite-alphabet $\bm s_1$, $\bm w$ is a Gaussian mixture with $|\C|^N$ components whose Fisher information admits no closed form, while a Gaussian surrogate yields only the worst case~\cite{stoica2011gauss}. Instead, we use Brown's identity~\cite{brown1982,payaro2009hessian}, which expresses this Fisher information with the minimum mean-square error (MMSE) in estimating an interfering data symbol from its noise-corrupted observation at the sensing receiver\rev{, which is efficiently computable}. Its complex vector form is stated as follows.

\begin{lemma}\label{lem:brown}
Let $\bm s_{\bm w}(\bm w)=\nabla_{\bm w^*}\ln p_{\bm w}(\bm w)$ denote the Wirtinger score of $\bm w=\bm A\bm s_1+\bm n$. Then, $\bm s_{\bm w}(\bm w)=-\sigma^{-2}\E[\bm n|\bm w]$ and
\begin{equation}\label{eq:brown}
\bm J_{\bm w}\triangleq\E[\bm s_{\bm w}\bm s_{\bm w}^H]=\frac{1}{\sigma^2}\left(\bm I-\frac{1}{\sigma^2}\E\big[\Cov(\bm A\bm s_1|\bm w)\big]\right).
\end{equation}
Under (A1) and (A4), $\E[\bm s_{\bm w}\bm s_{\bm w}^T]=\bm 0$.
\end{lemma}
\begin{proof}
Please refer to Appendix~\ref{app:lemma1}.
\end{proof}

\begin{theorem}\label{thm:main}
The FIM of $\bm\eta$ in~\eqref{eq:model} is given by
\begin{equation}\label{eq:fim}
[\bm F]_{ij}=2\,\Re\left\{\frac{\partial\bm\mu^H}{\partial\eta_i}\bm J_{\bm w}\frac{\partial\bm\mu}{\partial\eta_j}\right\}.
\end{equation}
Moreover, $\bm J_{\bm w}\succeq(\sigma^2\bm I+\bm A\bm A^H)^{-1}$ and the equality holds for Gaussian $\bm s_1$.
\end{theorem}
\begin{proof}
Please refer to Appendix~\ref{app:thm1}.
\end{proof}

To evaluate~\eqref{eq:brown}, consider the scalar channel $Y=\sqrt\gamma S_1+N$ with $S_1$ uniform on $\C$ and $N\sim\CN(0,1)$. Then, we define 
\begin{align}
    \mmse_{\C}(\gamma)&=\E[|S_1-\E[S_1|Y]|^2],
\end{align}
When the interference decouples into such scalar channels, Theorem~\ref{thm:main} reduces to the following closed form.

\begin{corollary}\label{cor:sep}
Let us define $\rho_{\C}(\gamma)=1-\gamma\,\mmse_{\C}(\gamma)$. If $\bm A$ has orthogonal columns, i.e., $\bm A=\bm Q\bm\Lambda$ with unitary $\bm Q$ and $\bm\Lambda=\mathrm{diag}(\lambda_1,\dots,\lambda_N)$, then
\begin{equation}\label{eq:sep}
\bm J_{\bm w}=\frac{1}{\sigma^2}\bm Q\,\mathrm{diag}\big(\rho_{\C}(\gamma_1),\dots,\rho_{\C}(\gamma_N)\big)\bm Q^H,\quad \gamma_i=\frac{|\lambda_i|^2}{\sigma^2}.
\end{equation}
If $\gamma_i=\gamma$ for all $i$, then $\bm F=\rho_{\C}(\gamma)\bm F_0$, where $\bm F_0$ is the interference-free FIM. Hence, the entire CRB matrix is inflated by the same factor, which we define as the CRB loss
\begin{equation}\label{eq:loss}
L_{\C}(\gamma)=\frac{1}{\rho_{\C}(\gamma)}=\frac{1}{1-\gamma\,\mmse_{\C}(\gamma)}.
\end{equation}
\end{corollary}
\begin{proof}
Please refer to Appendix~\ref{app:cor1}.
\end{proof}
Note that $\bm F$ is conditioned on the known $\bm s_0$ and marginalized over $\bm s_1$. Since $\bm F=\rho_{\mathcal S}(\gamma)\bm F_0$ holds for every realization of $\bm s_0$, $L_{\mathcal S}(\gamma)$ is identical whether or not the CRB is further averaged over $\bm s_0$.

\begin{proposition}\label{prop:props}
The CRB loss $L_{\C}(\gamma)$ satisfies the following properties:
\begin{enumerate}
\item[(i)] $1\le L_{\C}(\gamma)\le1+\gamma$, where the upper bound holds with equality for Gaussian $S_1$ and equals the loss of MF, i.e., treating interference as noise (TIN).
\item[(ii)] $\lim_{\gamma\to0}\frac{L_{\C}(\gamma)-1}{\gamma}=1$. Hence, at low INR, finite-alphabet interference is as harmful as Gaussian interference.
\item[(iii)] $L_{\C}(\gamma)\to1$ as $\gamma\to\infty$ for any finite $\C$.
\item[(iv)] $L_{\C}(\gamma)$ is maximized at a finite INR $\gamma^\star$.
\end{enumerate}
\end{proposition}
\begin{proof}
Please refer to Appendix~\ref{app:prop2}.
\end{proof}
\begin{remark}
Notably, Proposition~\ref{prop:props} reveals that, unlike the TIN loss $L_{\rev{\mathrm{TIN}}}(\gamma) = 1+\gamma$ growing without bound, the loss under finite-alphabet interference is non-monotonic: it follows the Gaussian loss at low INR, peaks at $\gamma^\star$, and vanishes at high INR, where the interfering symbols can be reliably detected and canceled as shown in Fig. \ref{fig:loss}.
\end{remark}

\subsection{Dependence of the FIM on the Interferer Waveform}\label{sec:wf}
By Theorem~\ref{thm:main}, the waveform of the interferer affects the FIM only through $\bm A=\bm H_{\mathrm I}\bm U_1$, and Corollary~\ref{cor:sep} yields a closed form when $\bm A$ has orthogonal columns. The following proposition gives a sufficient condition for waveform independence.

\begin{proposition}\label{prop:wf}
If the BS-BS channel is flat-fading, \rev{i.e.,} $\bm H_{\mathrm I}=h\bm I$, then $\bm J_{\bm w}=\rho_{\C}(\gamma)\bm I/\sigma^2$ for every unitary $\bm U_1$. Hence, $L_{\C}(\gamma)$ is identical for every interferer waveform (SC, CP-OFDM, OTFS, DFT-s-OFDM, AFDM, etc.), and $\bm F=\rho_{\C}(\gamma)\bm F_0$ holds for any waveform $\bm U_0$.
\end{proposition}
\begin{proof}
Since $\bm A=h\bm U_1$, Corollary~\ref{cor:sep} applies with $\bm Q=\bm U_1$ and $\lambda_i=h$, i.e., $\gamma_i=|h|^2/\sigma^2=\gamma$ for all $i$. Thus, $\bm J_{\bm w}=\rho_{\C}(\gamma)\bm I/\sigma^2$ does not depend on $\bm U_1$, and $\bm F=\rho_{\C}(\gamma)\bm F_0$ follows for any $\bm\mu(\bm\eta)$, i.e., any $\bm U_0$.
\end{proof}

Under a frequency-selective channel, $\bm H_{\mathrm I}=\rev{\bm W_N^H}\mathrm{diag}(H_k)\rev{\bm W_N}$, so that $\bm A$ has orthogonal columns for CP-OFDM ($\bm Q=\rev{\bm W_N^H}$, $\gamma_k=|H_k|^2/\sigma^2$), whose loss averages the scalar losses over the channel-induced INR spread. In contrast, $\bm A^H\bm A=\bm U_1^H\rev{\bm W_N^H}\mathrm{diag}(|H_k|^2)\rev{\bm W_N}\bm U_1$ is not diagonal for SC and OTFS, so that~\eqref{eq:brown} involves the joint MMSE attainable only by sequence-level processing. Since BS-BS links are typically LoS-dominated with a small delay spread, the flat-channel assumption of Proposition~\ref{prop:wf} is a reasonable approximation in practice.

\subsection{EM-Based Soft Interference Cancellation}\label{sec:em}
The exact FIM in Theorem~\ref{thm:main} provides an asymptotic benchmark for maximum-likelihood (ML) estimation under standard regularity conditions. To approach this benchmark, we employ an expectation-maximization (EM)-based receiver that iteratively estimates and suppresses the finite-alphabet interference\rev{~\cite{dempster1977em}}. Given the current estimate $\hat{\bm\eta}$, the E-step reduces to computing the posterior mean $\bar{\bm s_1}=\E[\bm s_1|\bm r-\bm\mu(\hat{\bm\eta})]$. In the separable case of Corollary~\ref{cor:sep}, it decouples into per-symbol computations on $\rev{\bm v}=\bm Q^H(\bm r-\bm\mu(\hat{\bm\eta}))$ as
\begin{equation}\label{eq:em}
\bar s_{1,i}=\frac{\sum_{c\in\C}c\,e^{-|\rev{v_i}-\lambda_ic|^2/\sigma^2}}{\sum_{c\in\C}e^{-|\rev{v_i}-\lambda_ic|^2/\sigma^2}},
\end{equation}
whereas the non-separable case requires the joint posterior over all $|\C|^N$ candidates of $\bm s_1$. Since the expected complete-data log-likelihood equals $-\|\bm r-\bm A\bar{\bm s_1}-\bm\mu(\bm\eta)\|^2/\sigma^2$ up to a constant, the M-step is the conventional interference-free estimator applied to the soft-cancelled signal $\bm r-\bm A\bar{\bm s_1}$, similar to soft interference cancellation as in iterative multiuser detection\rev{~\cite{wang1999turbo}}. In the separable case, each iteration requires only $N|\C|$ distance evaluations, without backhaul data sharing or decoding. As summarized in Algorithm~\ref{alg:em}, the receiver is initialized interference-first with $\hat{\bm\mu}=\bm 0$ in the first E-step, so that it remains robust when strong interference makes the initial MF estimate unreliable.

\begin{algorithm}[t]
\caption{EM-Based Soft Interference Cancellation}\label{alg:em}
\begin{algorithmic}[1]
\Require $\bm r$, own data $\bm s_0$, $\bm A=\bm Q\bm\Lambda$, $\C$, $\sigma^2$, and number of iterations $T$
\State Initialize $\hat{\bm\mu}\gets\bm 0$ (interference-first)
\For{$t=1,\dots,T$}
\State $\rev{\bm v}\gets\bm Q^H(\bm r-\hat{\bm\mu})$
\State \textbf{E-step:} compute $\bar s_{1,i}$, $i=1,\dots,N$, via~\eqref{eq:em}
\State \textbf{Soft cancellation:} $\tilde{\bm r}\gets\bm r-\bm A\bar{\bm s_1}$
\State \textbf{M-step:} $\hat{\bm\eta}\gets\arg\min_{\bm\eta}\|\tilde{\bm r}-\bm\mu(\bm\eta)\|^2$ (global search at $t=1$, local refinement for $t>1$)
\State $\hat{\bm\mu}\gets\bm\mu(\hat{\bm\eta})$
\EndFor
\Ensure $\hat{\bm\eta}$
\end{algorithmic}
\end{algorithm}

\section{Numerical Results}\label{sec:num}
Unless otherwise stated, both BSs employ CP-OFDM over a flat-fading BS-BS channel, for which Corollary~\ref{cor:sep} holds with $\gamma_i=\gamma$. The MMSE in~\eqref{eq:loss} has no closed form but is an expectation over the Gaussian noise, $\mmse_{\C}(\gamma)=\frac{1}{|\C|}\sum_{c\in\C}\E_N\big[|c-\hat s_1(\sqrt\gamma c+N)|^2\big]$ with $\hat s_1(y)=\E[S_1|Y=y]$ given in closed form by~\eqref{eq:em} \rev{with $\lambda_i=\sqrt\gamma$ and $\sigma^2=1$}. We evaluate it deterministically by Gauss-Hermite quadrature instead of Monte Carlo simulation.

\subsection{CRB Loss versus INR}
\begin{figure}[t]
\centering
\includegraphics[width=0.8\columnwidth]{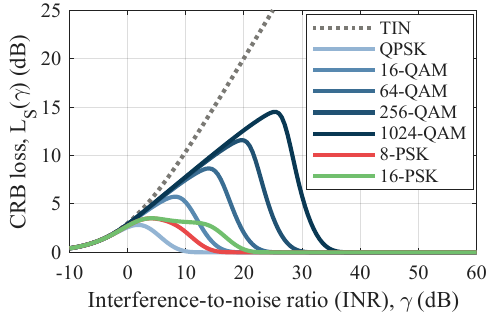}
\caption{CRB loss $L_{\C}(\gamma)$ in~\eqref{eq:loss} versus INR. The dotted line corresponds to Gaussian interference, which equals the loss of MF (TIN).}
\label{fig:loss}
\end{figure}

Fig.~\ref{fig:loss} evaluates the CRB loss \rev{in~\eqref{eq:loss} for various constellations}. All constellations follow the Gaussian curve at low INR, reach a peak, and then collapse to $0\dB$. The peak losses of QPSK, 16QAM, 64QAM, 256QAM, and 1024QAM are $2.8$, $5.7$, $8.7$, $11.6$, and $14.5\dB$ at $\gamma^\star=2$, $8$, $14$, $19.5$, and $25.5\dB$, respectively. That is, each fourfold increase of the QAM order raises the peak by approximately $2.9\dB$ and shifts it by approximately $6\dB$. PSK exhibits a lower peak of approximately $3.5\dB$ owing to its constant amplitude. For 64QAM at $\gamma=20\dB$, MF suffers a $20\dB$ CRB loss, whereas the optimal receiver can lose only $2\dB$.

\subsection{Receiver Performance}
\begin{figure}[t]
\centering
\includegraphics[width=0.8\columnwidth]{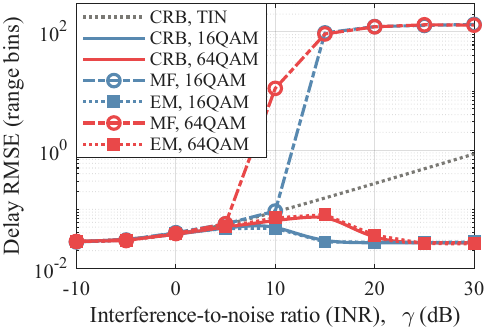}
\caption{Delay estimation RMSE versus INR under 16QAM and 64QAM.}
\label{fig:delay}
\end{figure}

\begin{figure*}[t]
\centering
\includegraphics[width=0.92\textwidth]{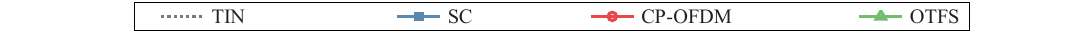}\\
\begin{minipage}{0.3\textwidth}\centering
\includegraphics[width=\linewidth]{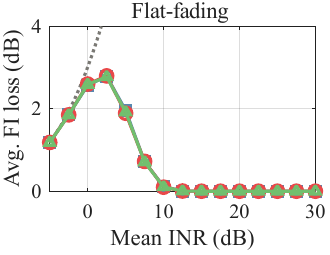}\\{\footnotesize (a)}
\end{minipage}\hfill
\begin{minipage}{0.3\textwidth}\centering
\includegraphics[width=\linewidth]{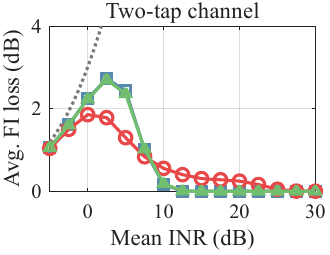}\\{\footnotesize (b)}
\end{minipage}\hfill
\begin{minipage}{0.3\textwidth}\centering
\includegraphics[width=\linewidth]{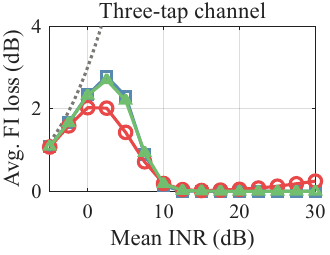}\\{\footnotesize (c)}
\end{minipage}
\caption{Average Fisher-information loss for a QPSK interferer over (a) a flat-fading channel, (b) a two-tap channel, and (c) a three-tap channel.}
\label{fig:wf}
\end{figure*}

Fig.~\ref{fig:delay} evaluates delay estimation with $N=256$ subcarriers and 8 OFDM symbols, where BS$_0$ transmits QPSK data with a per-sample echo SNR of $-10\dB$ and BS$_1$ transmits 16QAM or 64QAM data. The EM receiver runs 10 iterations with a FFT-based delay estimator in the M-step, and the RMSE is averaged over 1000 trials.
 Since Corollary~\ref{cor:sep} scales the entire FIM, the delay CRB is inflated by the same factor $L_{\C}(\gamma)$. The RMSE of MF follows the TIN bound, which \rev{equals} $\sqrt{1+\gamma}$ times the interference-free bound, and breaks down at $\gamma\approx10\dB$, where the integrated SINR falls below the detection threshold. In contrast, the EM receiver follows the finite-alphabet bound and enables sensing at INRs where TIN fails completely.

\subsection{Waveform Comparison}
Fig.~\ref{fig:wf} compares SC, CP-OFDM, and OTFS ($N_{\mathrm D}=2$) interferers with QPSK over a flat-fading channel and two frequency-selective channels (a two-tap channel with $\min_k|H_k|^2=0.023$ and a three-tap channel with $\min_k|H_k|^2=0.001$), where $\gamma$ denotes the mean INR. We evaluate~\eqref{eq:brown} with 1000 Monte Carlo samples of $\bm w$ per point.
 For a non-separable $\bm A$, this requires the joint posterior of $\bm s_1$ over $|\C|^N$ hypotheses, whose number grows exponentially with $N$. Therefore, we set $N=8$ with $4^8=65{,}536$ hypotheses. Moreover, since $\bm J_{\bm w}$ is no longer a scaled identity, the CRB loss depends on the echo derivative in~\eqref{eq:fim}. We thus provide the average Fisher-information (FI) loss $1/\rho_{\mathrm{eff}}$ with $\rho_{\mathrm{eff}}=\sigma^2\tr(\bm J_{\bm w})/N$, which reduces to $L_{\C}(\gamma)$ under Proposition~\ref{prop:wf}. Over the flat-fading channel in Fig.~\ref{fig:wf}(a), all waveforms coincide, confirming Proposition~\ref{prop:wf}. Over the frequency-selective channels, SC and OTFS with joint processing remain close to the flat curve, since each symbol spreads over all subcarriers and thus experiences the full channel energy. In contrast, CP-OFDM exhibits a lower peak but a residual tail, since deep-fade subcarriers remain in the harmful INR band as shown in Fig.~\ref{fig:wf}(b) and ~\ref{fig:wf}(c). 

\section{Conclusion}
In this paper, we have presented an estimation-theoretic framework for ISAC sensing under finite-alphabet inter-BS interference. For linear receivers, the impact of the inter-BS interference is invariant to the constellation and waveform. The exact FIM is scaled by $1-\gamma\,\mmse_{\C}(\gamma)$, so that the loss peaks at intermediate INR, vanishes at high INR, is waveform-independent for flat-fading and synchronized BS-BS links, and is \rev{approached} by EM-based soft cancellation. Hence, inter-BS coordination should avoid intermediate INRs, where increasing the interference power may even improve sensing.

\appendices

\section{Proof of Proposition~\ref{prop:cross}}\label{app:prop1}
Since $\E[\bm s_1\bm s_1^H]=\bm I_N$ and $\bm G$ is unitary, the second moment is
\begin{align}
\E[|\chi|^2]&=\E\big[\bm s_0^H\bm G^H\E[\bm s_1\bm s_1^H]\bm G\bm s_0\big]=\E\big[\|\bm s_0\|^2\big]=N.
\end{align}
For the fourth moment, let $\bm g=\bm G\bm s_0$ so that $\chi=\sum_ig_is_{1,i}^*$. For i.i.d.\ proper symbols with unit power and kurtosis $\mu_4$, the only non-zero fourth-order moments are $\E[s_i^*s_j^*s_ks_l]=\mu_4$ for $i=j=k=l$ and $1$ for $\{i,j\}=\{k,l\}$ with $i\neq j$, since $\E[s^2]=0$. Conditioning on $\bm g$ yields
\begin{equation}\label{eq:a2}
\E\big[|\chi|^4\,\big|\,\bm g\big]=2\|\bm g\|^4+(\mu_{4,s_1}-2)\sum_{i=1}^N|g_i|^4.
\end{equation}
Since $\|\bm g\|=\|\bm s_0\|$, the first term gives
\begin{align}
\E\big[\|\bm s_0\|^4\big]&=\sum_{i}\E[|s_{0,i}|^4]+\sum_{i\neq j}\E[|s_{0,i}|^2]\E[|s_{0,j}|^2]\nonumber\\
&=N^2+N(\mu_{4,s_0}-1).\label{eq:a3}
\end{align}
Applying~\eqref{eq:a2} again to $g_i=\sum_jG_{ij}s_{0,j}$ and using $\sum_j|G_{ij}|^2=1$ for each row of the unitary $\bm G$,
\begin{align}
\sum_{i=1}^N\E\big[|g_i|^4\big]&=\sum_{i=1}^N\Big(2+(\mu_{4,s_0}-2)\sum_j|G_{ij}|^4\Big)\nonumber\\
&=2N+(\mu_{4,s_0}-2)\|\bm G\|_4^4.\label{eq:a4}
\end{align}
Substituting~\eqref{eq:a3} and~\eqref{eq:a4} into~\eqref{eq:a2} yields~\eqref{eq:4th}. \rev{Finally, since each row of the unitary $\bm G$ has unit norm, $1/N\le\sum_j|G_{ij}|^4\le1$ with equality iff $|G_{ij}|^2=1/N$ and iff the row has a single non-zero entry, respectively, which gives $\|\bm G\|_4^4\in[1,N]$}, completing the proof.

\section{Proof of Lemma~\ref{lem:brown}}\label{app:lemma1}
The density of $\bm w$ is $p_{\bm w}(\bm w)=\E_{\bm s_1}[p_{\bm n}(\bm w-\bm A\bm s_1)]$ with $p_{\bm n}(\bm n)=(\pi\sigma^2)^{-N}e^{-\|\bm n\|^2/\sigma^2}$. Since $\nabla_{\bm w^*}p_{\bm n}(\bm w-\bm A\bm s_1)=-\sigma^{-2}(\bm w-\bm A\bm s_1)\,p_{\bm n}(\bm w-\bm A\bm s_1)$, Bayes' rule gives
\begin{align}
\bm s_{\bm w}(\bm w)=\frac{\nabla_{\bm w^*}p_{\bm w}(\bm w)}{p_{\bm w}(\bm w)}&=-\frac{\E_{\bm s_1}\big[(\bm w-\bm A\bm s_1)\,p_{\bm n}(\bm w-\bm A\bm s_1)\big]}{\sigma^2\,p_{\bm w}(\bm w)}\nonumber\\
&=-\sigma^{-2}\E[\bm n|\bm w].\label{eq:b1}
\end{align}
By the law of total covariance with $\E[\bm n]=\bm 0$, $\E_{\bm n}[\bm n\bm n^H]=\E_{\bm w}[\Cov(\bm n|\bm w)]+\E_{\bm w}\big[\E[\bm n|\bm w]\E[\bm n|\bm w]^H\big]$, and hence
\begin{equation}\label{eq:b2}
\bm J_{\bm w}=\frac{1}{\sigma^4}\E\big[\E[\bm n|\bm w]\E[\bm n|\bm w]^H\big]=\frac{1}{\sigma^4}\Big(\sigma^2\bm I-\E\big[\Cov(\bm n|\bm w)\big]\Big).
\end{equation}
Since $\bm n=\bm w-\bm A\bm s_1$ and $\bm w$ is given, $\Cov(\bm n|\bm w)=\Cov(\bm A\bm s_1|\bm w)$, and substituting this into~\eqref{eq:b2} yields~\eqref{eq:brown}.

Under (A1) and (A4), the circularity of $\bm n$ implies that $j\bm w$ has the same distribution as $\bm w$, satisfying $p_{\bm w}(j\bm w)=p_{\bm w}(\bm w)$. The Wirtinger chain rule with $\partial(j\bm w)^*/\partial\bm w^*=-j\bm I$ yields $\bm s_{\bm w}(j\bm w)=j\bm s_{\bm w}(\bm w)$, and therefore
\begin{equation}
\E[\bm s_{\bm w}\bm s_{\bm w}^T]=\E\big[\bm s_{\bm w}(j\bm w)\bm s_{\bm w}(j\bm w)^T\big]=-\E[\bm s_{\bm w}\bm s_{\bm w}^T],
\end{equation}
which implies $\E[\bm s_{\bm w}\bm s_{\bm w}^T]=\bm 0$. This completes the proof.

\section{Proof of Theorem~\ref{thm:main}}\label{app:thm1}
The log-likelihood is given by $\ln p_{\bm w}(\bm r-\bm\mu(\bm\eta))$. Let $\bm a_i=\partial\bm\mu/\partial\eta_i$. Since $\ln p_{\bm w}$ is real-valued, $\nabla_{\bm w}\ln p_{\bm w}=\bm s_{\bm w}^*$, and the chain rule gives
\begin{equation}\label{eq:c1}
\frac{\partial}{\partial\eta_i}\ln p_{\bm w}(\bm r-\bm\mu)=-\bm s_{\bm w}^H\bm a_i-\bm s_{\bm w}^T\bm a_i^*=-2\Re\{\bm s_{\bm w}^H\bm a_i\}.
\end{equation}
Using $\Re\{u\}\Re\{v\}=\frac12\Re\{uv^*\}+\frac12\Re\{uv\}$ with $u=\bm s_{\bm w}^H\bm a_i$ and $v=\bm s_{\bm w}^H\bm a_j$,
\begin{align}
[\bm F]_{ij}&=4\E\big[\Re\{\bm s_{\bm w}^H\bm a_i\}\Re\{\bm s_{\bm w}^H\bm a_j\}\big]\nonumber\\
&=2\Re\{\bm a_j^H\bm J_{\bm w}\bm a_i\}+2\Re\big\{\bm a_i^T\E[\bm s_{\bm w}^*\bm s_{\bm w}^H]\bm a_j\big\},\label{eq:c2}
\end{align}
where the second term vanishes by Lemma~\ref{lem:brown}, and $\Re\{\bm a_j^H\bm J_{\bm w}\bm a_i\}=\Re\{\bm a_i^H\bm J_{\bm w}\bm a_j\}$ since $\bm J_{\bm w}$ is Hermitian, which gives~\eqref{eq:fim}. For the bound, the MMSE error covariance is dominated by that of the LMMSE estimator, i.e., $\E[\Cov(\bm A\bm s_1|\bm w)]\preceq\bm B-\bm B(\bm B+\sigma^2\bm I)^{-1}\bm B$ with $\bm B=\bm A\bm A^H$, with equality for Gaussian $\bm s_1$. Substituting this into~\eqref{eq:brown} yields
\begin{equation}
\bm J_{\bm w}\succeq\frac{1}{\sigma^2}\big(\bm I-\bm B(\bm B+\sigma^2\bm I)^{-1}\big)=(\sigma^2\bm I+\bm A\bm A^H)^{-1}.
\end{equation}

\section{Proof of Corollary~\ref{cor:sep}}\label{app:cor1}
If $\bm A=\bm Q\bm\Lambda$, the sufficient statistic $\rev{\bm v}=\bm Q^H\bm w=\bm\Lambda\bm s_1+\bm Q^H\bm n$ consists of $N$ independent scalar channels $\rev{v_i}=\lambda_is_{1,i}+\tilde n_i$, since $\bm Q^H\bm n\sim\CN(\bm 0,\sigma^2\bm I)$ and the entries of $\bm s_1$ are i.i.d. Normalizing by $\sigma$ and noting that the MMSE is invariant to the known phase of $\lambda_i$, we have $\E[\Cov(\bm s_1|\bm w)]=\mathrm{diag}(\mmse_{\C}(\gamma_i))$, and thus
\begin{align}
\E\big[\Cov(\bm A\bm s_1|\bm w)\big]&=\bm Q\bm\Lambda\,\mathrm{diag}\big(\mmse_{\C}(\gamma_i)\big)\bm\Lambda^H\bm Q^H\nonumber\\
&=\sigma^2\bm Q\,\mathrm{diag}\big(\gamma_i\,\mmse_{\C}(\gamma_i)\big)\bm Q^H.
\end{align}
Substituting this into~\eqref{eq:brown} yields~\eqref{eq:sep}. If $\gamma_i=\gamma$, then $\bm J_{\bm w}=\rho_{\C}(\gamma)\bm I/\sigma^2$, so that~\eqref{eq:fim} becomes $\rho_{\C}(\gamma)$ times the interference-free FIM $\bm F_0$~\cite{kay1993}, and $\bm F^{-1}=\bm F_0^{-1}/\rho_{\C}(\gamma)$, completing the proof.

\section{Proof of Proposition~\ref{prop:props}}\label{app:prop2}

(i) Consider the linear estimator $\hat S_{1,\rm L}=\frac{\sqrt{\gamma}}{1+\gamma}Y$, whose mean-square error is $1/(1+\gamma)$. Since the MMSE estimator minimizes the mean-square error, we have $0\le\mmse_{\C}(\gamma)\le 1/(1+\gamma)$. Substituting these bounds into~\eqref{eq:loss} yields $1\le L_{\C}(\gamma)\le 1+\gamma$. The upper bound is attained for Gaussian $S_1$, for which the conditional mean estimator is linear.

(ii) From~\eqref{eq:loss}, we have
\begin{equation}
\frac{L_{\C}(\gamma)-1}{\gamma}=\frac{\mmse_{\C}(\gamma)}{1-\gamma\mmse_{\C}(\gamma)}.
\end{equation}
As $\gamma\to0$, the observation $Y=\sqrt{\gamma}S_1+N$ becomes independent of $S_1$. Consequently, $\E[S_1|Y]$ converges to $\E[S_1]=0$, implying $\mmse_{\C}(\gamma)\to\E[|S_1|^2]=1$. Since $\gamma\mmse_{\C}(\gamma)\to0$, it follows that $\lim_{\gamma\to0}(L_{\C}(\gamma)-1)/\gamma=1$.

(iii) Let $\hat S_{1,\rm MD}$ denote the minimum-distance detector, and let $d_{\min}$ and $d_{\max}$ be the minimum and maximum distances between distinct constellation points, respectively. Since the MMSE estimator minimizes the mean-square error, we have
\begin{equation}
\mmse_{\C}(\gamma)\le\E[|S_1-\hat S_{1,\rm MD}|^2]\le d_{\max}^2 P_e(\gamma),
\end{equation}
where $P_e(\gamma)=\Pr(\hat S_{1,\rm MD}\neq S_1)$. For an $M$-ary constellation in complex AWGN, the union bound and $Q(x)\le\frac12e^{-x^2/2}$ give
\begin{equation}
P_e(\gamma)\le(M-1)Q\left(\sqrt{\frac{\gamma d_{\min}^2}{2}}\right)\le\frac{M-1}{2}e^{-\gamma d_{\min}^2/4}.
\end{equation}
Combining these inequalities yields
\begin{equation}
0\le\gamma\mmse_{\C}(\gamma)\le\frac{(M-1)d_{\max}^2}{2}\gamma e^{-\gamma d_{\min}^2/4}.
\end{equation}
Since $d_{\min}>0$, the right-hand side converges to zero as $\gamma\to\infty$. Thus, $\gamma\mmse_{\C}(\gamma)\to0$, and~\eqref{eq:loss} gives $L_{\C}(\gamma)\to1$.

(iv) $L_{\C}(\gamma)$ is continuous and approaches unity as $\gamma\to0$ and $\gamma\to\infty$, as established in (ii) and (iii)\rev{, and exceeds unity for $\gamma>0$ since $\mmse_{\C}(\gamma)>0$}. Therefore, there exist $0<a<b<\infty$ such that the global maximum is attained within $[a,b]$. By continuity, $L_{\C}(\gamma)$ attains its maximum at some finite $\gamma^\star\in(0,\infty)$, completing the proof.

\bibliographystyle{IEEEtran}
\bibliography{refs}

\end{document}